\documentclass[a4paper,12pt]{article}

\usepackage{graphicx}
\usepackage{epsfig}
\usepackage{amsfonts,amsmath}
\usepackage{amssymb}
\usepackage{epsf}
\usepackage{hyperref}
\usepackage{authblk}

\usepackage{color}
\usepackage[normalem]{ulem}
\usepackage[dvipsnames]{xcolor}
\usepackage{cite}

\title{Stationary Einstein-vector-Gauss-Bonnet black holes}

\author[1]{Burkhard Kleihaus \thanks{\href{mailto:b.kleihaus@uni-oldenburg.de}{b.kleihaus@uni-oldenburg.de}}}
\author[1]{Jutta Kunz\thanks{\href{mailto:jutta.kunz@uni-oldenburg.de}{jutta.kunz@uni-oldenburg.de}}}

\affil[1]{Institut f\"ur  Physik, Universit\"at Oldenburg, Postfach 2503,
D-26111 Oldenburg, Germany}

\date{\today}

\begin{document}

\maketitle

\begin{abstract}
We study spontaneously vectorized black holes in Einstein-vector-Gauss-Bonnet theory with a quadratic coupling function.
Besides the static, spherically symmetric black holes carrying an electric charge, there are uncharged static, axially symmetric black holes that possess a magnetic dipole moment.
Both types possess radial excitations.
The magnetic black holes are prolate.
They are hotter than the Schwarzschild black holes and possess lower free energy.
The domain of existence of the rotating vectorized black holes is bounded by the Kerr black holes, the spherically and axially symmetric static black holes, and the critical solutions.

\end{abstract}

\section{Introduction}

General Relativity (GR) may be viewed as the leading term in a systematic low-energy expansion of gravitational dynamics.
In such an effective field theory (EFT) approach the expansion is performed in powers of derivatives and curvature \cite{Donoghue:1995cz,Burgess:2003jk}.
An important direction in recent years has been to augment the gravitational EFT with additional light scalar or vector degrees of freedom.
Such modified gravity theories may be viewed as motivated by dark energy, dark matter, or beyond standard model physics \cite{Berti:2015itd,CANTATA:2021ktz,Faraoni:2010pgm} .

Among the scalar–-tensor EFTs, the Horndeski theories stand out since they give rise to second-order field equations \cite{Horndeski:1974wa,Charmousis:2011bf,Kobayashi:2011nu}.
Moreover, they allow for different types of scalarization of compact objects \cite{Doneva:2022ewd}.
Spontaneous scalarization was originally observed in scalar-tensor theory for sufficiently compact neutron stars, where it is matter-induced \cite{Damour:1993hw}. 
In contrast, spontaneous scalarization of the Schwarzschild and Kerr black holes relies on the presence of higher curvature terms in the EFTs, coupled to the scalar field. 

In Einstein-scalar-Gauss-Bonnet (EsGB) theories the higher curvature terms enter in the form of a topological invariant, the Gauss-Bonnet (GB) term.
The choice of coupling function is then decisive for the properties of the scalarized black holes that arise.
While spontaneous scalarization is typically associated with a tachyonic instability of GR black holes \cite{Doneva:2017bvd,Silva:2017uqg,Antoniou:2017acq,Cunha:2019dwb,Collodel:2019kkx,Dima:2020yac,Hod:2020jjy,Doneva:2020nbb,Herdeiro:2020wei,Berti:2020kgk}, in non-linear scalarization no such instability occurs \cite{Doneva:2021tvn,Blazquez-Salcedo:2022omw,Doneva:2022yqu,Lai:2023gwe}.
Scalarized black holes also arise in dilatonic and shift-symmetric  EsGB theories, which no longer allow for GR black holes \cite{Kanti:1995vq,Torii:1996yi,Pani:2009wy,Kleihaus:2011tg,Sotiriou:2014pfa}.

Generalized Proca theories are EFTs with a vector degree of freedom that give rise to second-order field equations \cite{Horndeski:1976gi,Tasinato:2014eka,Heisenberg:2014rta,Tasinato:2014mia}.
Such vector-tensor EFTs can also possess black hole solutions as studied, for instance, in \cite{Chagoya:2016aar,Fan:2016jnz,Babichev:2017rti,Chagoya:2017fyl,Heisenberg:2017xda,Heisenberg:2017hwb,Verbin:2020fzk,Oliveira:2020dru,Barton:2021wfj,Minamitsuji:2024ygi,Charmousis:2025jpx,Eichhorn:2025pgy,Konoplya:2025uiq,Lutfuoglu:2025qkt,Konoplya:2025bte,Fernandes:2026rjs}. 
Analogous to spontaneous scalarization, in such vector EFTs the phenomenon of spontaneous vectorization of black holes may occur  \cite{Ramazanoglu:2017xbl,Ramazanoglu:2018tig,Ramazanoglu:2019gbz,Ramazanoglu:2019jrr}.
In Einstein-vector-Gauss-Bonnet (EvGB) theories a vector field is coupled to the GB term with a suitable coupling function.
The resulting spontaneously vectorized black holes thus carry Proca hair \cite{Barton:2021wfj}.

The static, spherically symmetric, spontaneously vectorized black holes in EvGB theory carry an electric charge \cite{Barton:2021wfj}.
These electric black holes bifurcate from the Schwarzschild solutions at a certain value of the GB coupling constant $\lambda$, where the tachyonic instability arises and continue to exist for arbitrarily large values of $\lambda$.
Here we show that EvGB theory features a further tachyonic instability at a lower value of the coupling $\lambda$ where static vectorized black holes arise.
These carry a magnetic dipole moment and possess only axial symmetry.
Subsequently, we add rotation to both types of black holes and show that their domains of existence merge at sufficiently rapid rotation.

The paper is organized as follows:
Section 2 provides the theoretical setting. 
It specifies the action and gives the equations of motion and the Ans\"atze for the metric and the vector potential together with the boundary conditions.
It also specifies the expressions for the global charges, the magnetic dipole moment, the horizon properties, and some relevant thermodynamic properties.
The results are presented in section 3.
First, both types of static solutions and their properties are discussed together with a perturbative approach to their radial excitations.
Then the rotating vectorized black holes are presented.
We conclude in section 4.

\section{Theoretical setting}

\subsection{Action and equations of motion}

We consider the effective EvGB action 
\begin{eqnarray}  
S=\frac{1}{16 \pi}\int \left[R - F_{\mu\nu}F^{\mu\nu}  
 + \lambda A_\mu A^\mu R^2_{\rm GB}
 \right] \sqrt{-g} d^4x  \ ,
\label{act}
\end{eqnarray} 
with  curvature scalar $R$, Gauss-Bonnet invariant
\begin{eqnarray} 
R^2_{\rm GB} = R_{\mu\nu\rho\sigma} R^{\mu\nu\rho\sigma}
- 4 R_{\mu\nu} R^{\mu\nu} + R^2 \ , 
\end{eqnarray} 
and field strength tensor $F_{\mu\nu}$ of the vector field $A_\mu$.
The vector field is coupled to the Gauss-Bonnet term via the coupling function 
%$F(A_\mu A^\mu)=\lambda A_\mu A^\mu$, 
$\lambda A_\mu A^\mu$
with coupling parameter $\lambda$.

Note that the Gauss-Bonnet invariant $R^2_{\rm GB}$ itself is topological in four dimensions, but it yields nontrivial contributions to the equations of motion 
when coupled to the vector field $A_\mu$. 

The Proca equations and the Einstein equations are obtained by variation of the action (\ref{act}) with respect to the vector field and the metric
\begin{equation}
\nabla_\mu F^{\mu\nu} = -\frac{\lambda}{2} A^\nu R^2_{\rm GB} \ ,
\label{scleq}
\end{equation}
\begin{equation}
G_{\mu\nu}  = \frac{1}{2}T^{({\rm eff})}_{\mu\nu} \ , 
\label{Einsteq}
\end{equation}
with Einstein tensor $G_{\mu\nu}$. 
The effective stress-energy tensor $T^{({\rm eff})}_{\mu\nu}$ is denoted by
\begin{equation}
%T^{({\rm eff})}_{\mu\nu} = T^{(A)}_{\mu\nu} +  T^{(GB)}_{\mu\nu} \ ,
T^{({\rm eff})}_{\mu\nu} = T^{(A)}_{\mu\nu} -2 T^{(GB)}_{\mu\nu} \ ,
\label{teff}
\end{equation}
which contains a term from the field strength tensor
\begin{equation}
T^{(A)}_{\mu\nu} = 4 F_\mu^{\phantom{\mu}\kappa} F_{\nu\kappa} 
		    -g_{\mu\nu} \left(F_{\rho\kappa}F^{\rho\kappa}\right)\ , 
\label{tphi}
\end{equation}
and a term from the GB invariant $R^2_{\rm GB}$
\begin{equation}
T^{(GB)}_{\mu\nu} =
\frac{\lambda}{2}\left(g_{\rho\mu} g_{\beta\nu}+g_{\beta\mu} g_{\rho\nu}\right)
\eta^{\kappa\beta\alpha\beta}\tilde{R}^{\rho\gamma}_{\phantom{\rho\gamma}\alpha\beta}
\nabla_\gamma \nabla_\kappa \left(A_\mu A^\mu\right) + 
\lambda A_\mu A_\nu\, R^2_{\rm GB},
\label{teffi}
\end{equation}
where
$\tilde{R}^{\rho\gamma}_{\phantom{\rho\gamma}\alpha\beta}=\eta^{\rho\gamma\sigma\tau}
R_{\sigma\tau\alpha\beta}$ and $\eta^{\rho\gamma\sigma\tau}= 
\epsilon^{\rho\gamma\sigma\tau}/\sqrt{-g}$. Note that the last term results from the dependence of the coupling function on the metric. 

Here we restrict to a massless vector field. 
However, we found in a preliminary study that stationary rotating and non-rotating black hole solutions with massive vector field exist.

To obtain stationary, rotating, axially symmetric black holes, we employ isotropic coordinates for the line element
\begin{equation}
ds^2 = -f_0 dt^2 + f_1\left[ f_2 \left(dr^2+r^2 d\theta^2\right) 
                            +r^2\sin^2\theta\left( d\varphi-\omega dt\right)^2 \right]\ ,
\label{met}
\end{equation}
and we assume for the vector field the form
\begin{equation}
A_\mu dx^\mu = A_t dt + A_r dr + A_\theta d\theta + A_\varphi d\varphi \ .
\label{vec}
\end{equation}
All functions depend only on the radial coordinate $r$ and the polar angle $\theta$. 

Inserting the above ansatz (\ref{met})-(\ref{vec}) into the EvGB equations yields four partial differential equations (PDEs) for the vector field functions and six PDEs for the metric functions.

We note that it is consistent to set $A_r(r,\theta)=0$ and $A_\theta(r,\theta)=0$.
As a consequence the Lorentz condition $\nabla_\mu A^\mu =0$ is satisfied.

For convenience, we parametrize the remaining vector field components as
\begin{equation}
A_t= V -\omega H \sin\theta \ , \ \ \  A_\varphi = H \sin\theta \  ,
\label{repA}
\end{equation}
which yields for the coupling function
\begin{equation}
\lambda A_\mu A^\mu = \lambda \left(\frac{H^2}{r^2 f_1} - \frac{V^2}{f_0} \right) \ .
\label{coupA}
\end{equation}
Although we use terms like ``electric charge" and ``magnetic dipole moment", the vector field can not be identified with the electromagnetic gauge potential, since the coupling function $\lambda A_\mu A^\mu$ breaks gauge invariance.

% Inspection of the field equations reveals an invariance under the scaling transformation
% \begin{equation}
% r \to \chi r \ , \ \ \    t \to \chi t \ , \ \ \  \lambda \to \chi^2 \lambda \  , 
% \ \ \  \chi > 0 \ .
% \label{scalinvar}
% \end{equation}

\subsection{Boundary Conditions}

We introduce the compactified coordinate $x=1-r_{\rm H}/r$.
Thus the horizon $r=r_{\rm H}$ and spatial infinity are mapped to $x=0$ respectively $x=1$.

In order to extract the double zeros at the horizon we re-define the functions 
\begin{equation}
f_0 \to  x^2 f_{0}\ , \ \ \
V   \to  x^2 V  \ .
\label{f0vrh}
\end{equation}
The boundary conditions at the horizon result from regularity,
\begin{equation}
  f_0-\partial_x f_0 = 0 \, , \ \ \ 
2 f_1+\partial_x f_1 = 0 \, , \ \ \  
\partial_x f_2 = 0 \, , \ \ \  
\omega = \omega_{\rm H} \, , \ \ \ 
 V-\partial_x V = 0 \, , \ \ \   
\partial_x H = 0 \, .  
\label{BCrh}
\end{equation}
At spatial infinity, we require Minkowski spacetime and vacuum 
\begin{equation}
f_i = 1 \, , \ \ \ i=0,1,2 \, \ \ \ 
\omega =0 \ ,  \ \ \
V = 0 \, , \ \ \  H = 0 \, . 
\label{BCinf}
\end{equation}
On the symmetry axis, elementary flatness and regularity require
\begin{equation}
\partial_\theta f_0 = 0 \, , \  
\partial_\theta f_1 = 0 \, , \  
f_2 = 1 \, , \  
\partial_\theta \omega = 0 \, , \  
\partial_\theta V = 0 \, , \  
H = 0 \, . \ 
\label{BCz}
\end{equation}
Reflection symmetry with respect to the equatorial plane yields the boundary conditions
\begin{equation}
\partial_\theta f_i = 0 \, , \ \ \ i=0,1,2 \, \ ,  \ \ \
\partial_\theta \omega  = 0 \, , \  
\partial_\theta V = 0 \, , \  
\partial_\theta H = 0 \, .
\label{BCeq}
\end{equation}

\subsection{Physical Properties}

Mass $M$, angular momentum $J$, electric charge $Q$ and magnetic dipole moment $\mu$ can be obtained from the asymptotic behavior of the metric, respectively the vector field
\begin{equation}
x^2 f_0 \to 1-\frac{2 M}{r} \, , \ \ \ 
\omega  \to \frac{2 J}{r^3} \, , \ \ \
A_t  \to \frac{Q}{r} \, , \ \ \
A_\varphi  \to -\mu\frac{\sin^2\theta}{r} \, .
\label{MJinf}
\end{equation}

In the numerical construction we considered only black holes with non-negative $J$, $Q$ and $\mu$. 
Solutions with negative $J$, $Q$ or $\mu$ can be obtained by employing discrete symmetries of the Einstein and field equations.
Changing the sign of $J$ requires a change of sign in $\omega$.
In order not to change the relative sign of $V$ and $\omega H$ in Eq.~(\ref{repA}) we can either change the sign of $V$ or the sign
of $H$. 
The former yields a change of sign of the electric charge $Q$, while the magnetic dipole moment $\mu$ does not change sign.
The latter on the other hand leads to a change of sign of the dipole moment $\mu$ while the electric charge $Q$ does not change sign.
Consequently, the magnetic dipole moment always has the same sign as the product $Q J$. 
Note that for Kerr-Newman black holes the magnetic dipole moment $\mu_{\rm KN}= Q_{\rm KN} J_{\rm KN} / M_{\rm KN}$ has the same property.

Horizon area $A_{\rm H}$, entropy $S_{\rm H}$, Hawking temperature $T_{\rm H}$, polar radius $R_p$ and equatorial radius $R_e$ can be obtained from the horizon.

We define the metric of a spatial cross-section of the horizon $\Sigma_{\rm H}$
\begin{eqnarray}
\label{horizon-metric}
d\Sigma^2_{\rm H}=h_{ij} dx^i dx^j=
r_{\rm H}^2 f_1(r_{\rm H},\theta ) 
\left(  f_2(r_{\rm H},\theta ) d\theta^2 + \sin^2\theta d\varphi^2\right) .
\end{eqnarray}

This yields the horizon area of the black holes
\begin{eqnarray}
\label{AH}
A_{\rm H}=4\pi r_{\rm H}^2 \int_0^{\pi/2} f_1\sqrt{f_2}\sin\theta d\theta \, .
\end{eqnarray}
For vectorized black holes, the entropy consists of the GR term $A_{\rm H}/4$ and a contribution from the GB invariant \cite{Wald:1984rg,Lee:1990nz,Wald:1993nt,Iyer:1994ys,Hajian:2015xlp,Ghodrati:2016vvf,Hajian:2020dcq}
\begin{eqnarray}
\label{S-Noether} 
S_{\rm H}=
\frac{1}{4}\int_{\Sigma_{\rm H}} 
\sqrt{h}\left[1+ 2 \lambda A_\mu A^\mu \tilde R \right] d^{2}x
= \pi r_{\rm H}^2\int_0^{\pi/2}
\left[1+ 2 \lambda \frac{H^2}{r^2 f_1} \tilde R \right] f_1\sqrt{f_2}\sin\theta d\theta \, ,
\end{eqnarray} 
where $h$ denotes the determinant of the horizon metric, and $\tilde R$ is the horizon curvature.

The Hawking temperature $T_{\rm H}$ is related to the surface gravity $\kappa$, i.~e.~$T_{\rm H}={\kappa}/({2\pi})$, where $\kappa^2=-\frac{1}{2}(\nabla_a \chi_b)(\nabla^a \chi^b)|_{r_{\rm H}}$ is determined from the Killing vector field $\chi= \partial_t + \Omega_H \partial_\varphi$. 
Substitution of the metric yields

\begin{eqnarray}
\label{TH}
T_{\rm H}=\frac{1}{2 \pi r_{\rm H}} \sqrt{\frac{f_0}{f_1 f_2}} \, .
\end{eqnarray}
The free energy $F$ is given by
\begin{eqnarray}
\label{F}
F= M - T_H S\, .
\end{eqnarray}

The polar radius $R_p$ and equatorial radius $R_e$ are obtained as

\begin{eqnarray}
\label{RpRe}
R_p=  r_{\rm H} \frac{2}{\pi} \int_0^{\pi/2}\sqrt{f_1 f_2} d\theta \, , \ \ \ 
R_e =  r_{\rm H} \left. \sqrt{f_1} \right|_{\theta=\pi/2} \, ,
\end{eqnarray}
respectively.

%\end{document}

\section{Results}

\subsection{Numerics}

We put the Einstein equations in the form $E_\mu^{\ \nu}=0$ and solve the four PDEs 
$E_t^{\ t}=0$, $E_r^{\ r}+E_\theta^{\ \theta}=0$,
$E_\varphi^{\ \varphi}=0$, and $E_\varphi^{\ t}=0$.
The isotropic horizon coordinate $r_{\rm H}$ is kept fixed, while the parameters $\omega_{\rm H}$ and $\lambda$ are varied.
For the numerical construction of the solutions we use the finite difference method CADSOL and a pseudo spectral method as well.
Both methods use Newton-Raphson iterations to compute the numerical solutions from an initial configuration.
Critical solutions are reached when the Newton-Raphson iterations diverge.
For CADSOL we chose a typical grid size $121\times 51$ and order of consistency $6$.
The typical numerical errors are of the order $10^{-6}$.  
The pseudo spectral method is less time consuming. 
Typically the number of modes are $32\times 16$.
Both methods yield the (almost) same values for the critical parameters, e.~g.~for $\omega_{\rm H} = 0.05$ we find $\lambda_{\rm cr}=21.172 $ for the finite difference method and  $\lambda_{\rm cr}=21.173 $ for the spectral method.

As further checks on the numerical solutions we evaluated the Komar mass $M_K$ and the Komar angular momentum $J_K$ and compared them with the ADM mass $M$ and the angular momentum $J$ extracted from the asymptotic behaviour of the metric. 
Along the bifurcation line the relative deviations are very small.
Along the critical line we find $|M/M_K-1|\lesssim 10^{-3}$, while $|J/J_K-1|$ is typically of order $10^{-3}$ and increases up to the percent level only near the upper end of the critical boundary, where the numerical calculations become more demanding. 
The larger deviations in the angular momentum are expected, since $J$ is extracted from the asymptotic expansion of $\omega$ at order $1/r^3$, whereas $M$ is obtained from the $1/r$ coefficient of $   g_{tt}$. 
This makes the extraction of $J$ more sensitive to the numerical accuracy in the asymptotic region.
We also considered the identity 
\begin{equation}
\int F_{\mu\nu} F^{\mu\nu} \sqrt{-g} d^3x = 
\int \lambda A_\mu A^\mu R^2_{\rm GB} \sqrt{-g} d^3x \ ,
\label{S_FGB}
\end{equation}
which follows from the vector field equation (\ref{scleq}). 
We found deviations of the order $10^{-6}$, except for
critical solutions for large values of $\lambda$, where the 
deviation is of the order $10^{-4}$.

\subsection{Static black holes}

\begin{figure}[t!]
\begin{center}
%\vspace{-0.5cm}
\mbox{
(a)\hspace*{-0.5cm}\includegraphics[height=.25\textheight, angle =0]{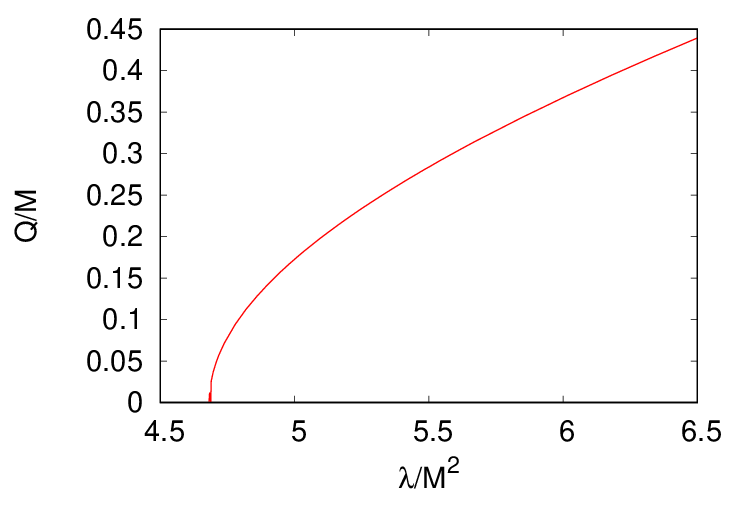}
(b)\hspace*{-0.5cm}\includegraphics[height=.25\textheight, angle =0]{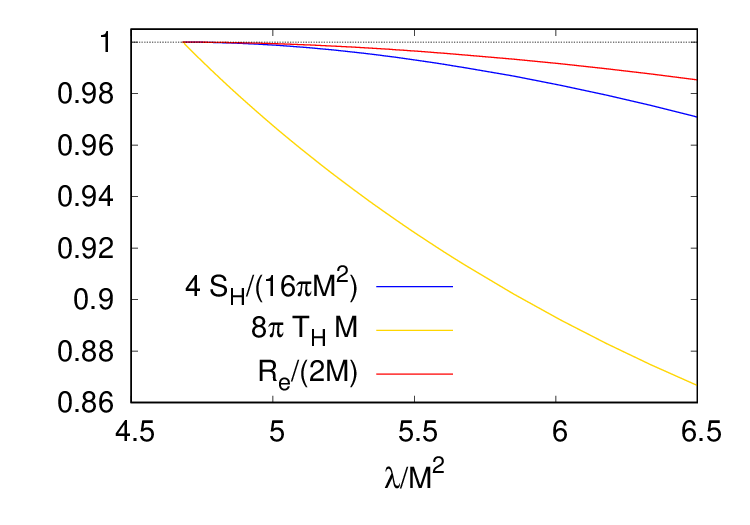}
}
\end{center}
\vspace{-0.5cm}
\caption{Static spherically symmetric EvGB black holes:
(a) scaled electric charge $Q/M$ vs scaled coupling $\lambda/M^2$; 
(b) scaled entropy $4 S_{\rm H}/(16\pi M^2)$, scaled Hawking temperature $8 \pi T_{\rm H} M$ and scaled equatorial horizon radius $R_e/(2M)$ vs scaled coupling.
}
\label{fig1}
\end{figure}

\begin{figure}[t!]
\begin{center}
%\vspace{-0.5cm}
\mbox{
\hspace*{-0.5cm}\includegraphics[height=.25\textheight, angle =0]{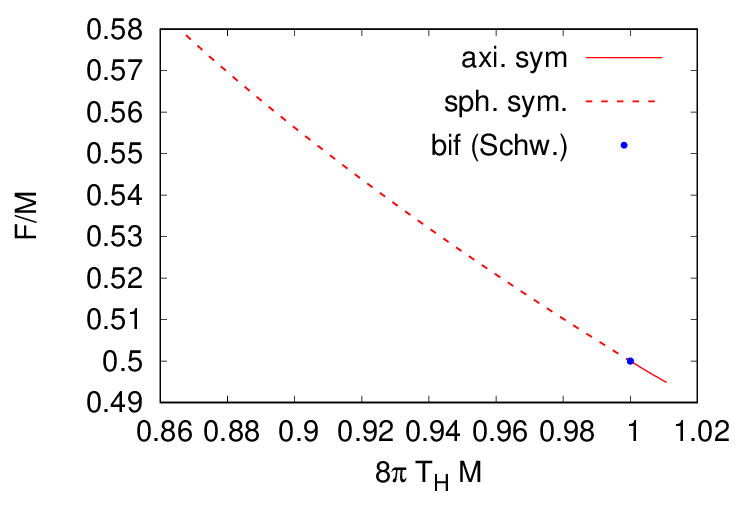}
}
\end{center}
\vspace{-0.5cm}
\caption{Static spherically and axially symmetric EvGB black holes:
Scaled free energy $F/M$ vs scaled Hawking temperature $8 \pi T_{\rm H} M$. 
}
\label{fig2}
\end{figure}

Static spherically symmetric EvGB black holes have been considered before in \cite{Barton:2021wfj}.
For these electric solutions, the $A_\varphi = g_{t\varphi} =0$, $f_2=1$ and $f_0$, $f_1$, and $V$ are functions of $r$ only. 
They are characterized by vanishing magnetic dipole moment.
Their properties are summarized in Fig.~\ref{fig1} in terms of dimensionless quantities.
In Fig.~\ref{fig1}(a) we show the scaled electric charge $Q/M$ versus the scaled coupling parameter $\lambda/M^2$.
These black holes bifurcate with the Schwarzschild black holes at the value of the coupling $(\lambda/M^2)_{\rm bif}=4.682$, where a tachyonic instability of the Schwarzschild black holes arises.
They then exist for arbitrarily large coupling \cite{Barton:2021wfj}.

We exhibit the scaled equatorial horizon radius $R_e/(2M)$, the scaled entropy $4 S_{\rm H}/(16\pi M^2)$, and the Hawking temperature $8 \pi T_{\rm H} M$ versus the scaled coupling parameter $\lambda/M^2$ in Fig.~\ref{fig1}(b). 
While the Schwarzschild values of these quantities are all equal to one, the corresponding values of the vectorized black holes decrease monotonically from the bifurcation point with increasing scaled coupling.
Thus, since their entropy is lower than the entropy of the Schwarzschild black holes, they are thermodynamically disfavored.
The same conclusion holds when their free energy $F$ is considered versus the temperature, since it is higher than the Schwarzschild free energy, as seen in Fig.~\ref{fig2}.

\begin{figure}[t!]
\begin{center}
%\vspace{-0.5cm}
\mbox{
(a)\hspace*{-0.5cm}\includegraphics[height=.25\textheight, angle =0]{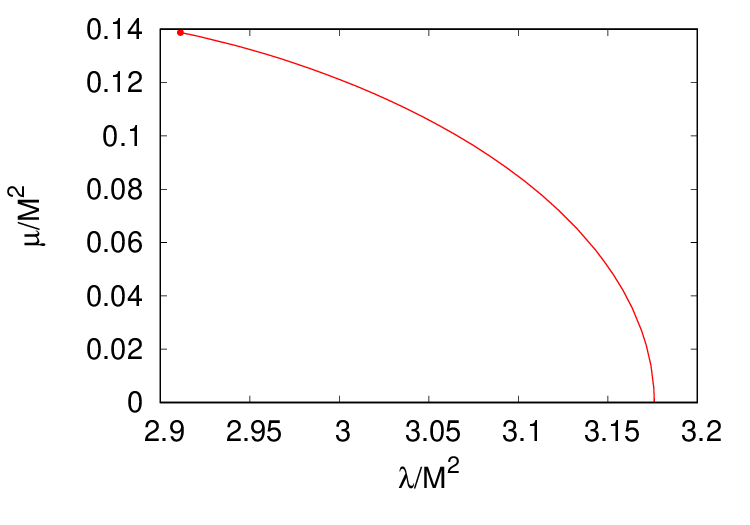}
(b)\hspace*{-0.5cm}\includegraphics[height=.25\textheight, angle =0]{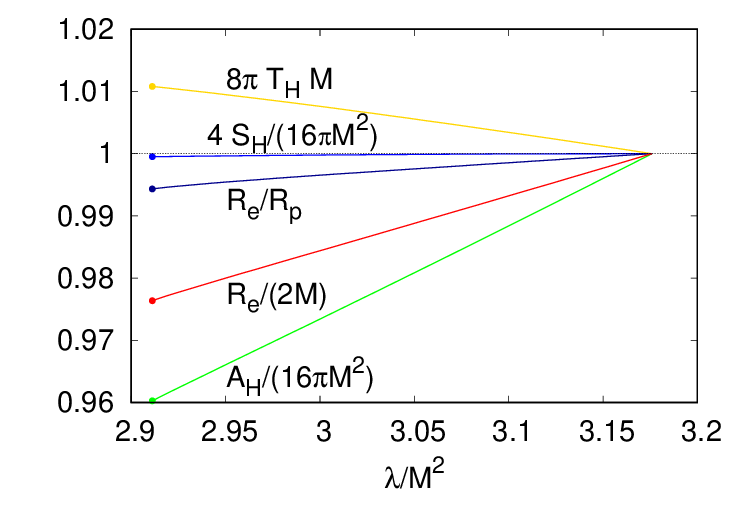}
}
\end{center}
\vspace{-0.5cm}
\caption{Static axially symmetric EvGB black holes:
(a) scaled magnetic dipole moment $\mu/M^2$ vs scaled coupling $\lambda/M^2$; 
(b) scaled entropy $4 S_{\rm H}/(16\pi M^2)$, scaled Hawking temperature $8 \pi T_{\rm H} M$, equatorial-to-polar horizon radius ratio $R_e/R_p$, scaled equatorial horizon radius $R_e/(2M)$, and scaled horizon area $A_{\rm H}/(16\pi M^2)$ vs scaled coupling.
}
\label{fig3}
\end{figure}

Static axially symmetric GR black holes are excluded by Israel's theorem \cite{Israel:1967wq}.
In EvGB theories they may exist, however, just as in EsGB theories.
For such magnetic EvGB black holes $A_t = g_{t\varphi} =0$, and the functions $f_0$, $f_1$, $f_2$ and $H$ depend on $r$ and $\theta$. 
They are characterized by vanishing electric charge.
The properties of these axially symmetric vectorized black holes are summarized in Fig.~\ref{fig3}.
Here the scaled magnetic dipole moment $\mu/M^2$ is shown versus the scaled coupling parameter $\lambda/M^2$ in Fig.~\ref{fig3}(a).
These solutions exist only over a limited range of the coupling parameter $(\lambda/M^2)_{\rm cr} < \lambda/M^2 \leq (\lambda/M^2)_{\rm bif}$.

The magnetic black holes bifurcate with the Schwarzschild black hole at $(\lambda/M^2)_{\rm bif} = 3.176$ and end at a critical solution at $(\lambda/M^2)_{\rm cr}=2.912$.
Here presumably some discriminant vanishes at the horizon\footnote{This would be analogous to scalarized EsGB black holes\cite{Collodel:2019kkx,Berti:2020kgk}, 
where we also do not observe a strong rise of curvature invariants.}. 
We note that, in contrast to the spherically symmetric branch of black holes, this axially symmetric branch tends to smaller values of the coupling.
This suggests that it might be physically preferred over the Schwarzschild branch.

The scaled entropy $4S_{\rm H}/(16\pi M^2)$ of these black holes, on the other hand, is almost degenerate with the Schwarzschild one, as seen in Fig.~\ref{fig3}(b),
but it remains slightly below the Schwarzschild value. 
Thus, if the entropy at fixed ADM mass is used as the criterion, the magnetic vectorized black holes are still thermodynamically disfavored with respect to the Schwarzschild black holes. 
The near degeneracy is nevertheless noteworthy, since the lower area contribution is almost exactly compensated by the additional Gauss-Bonnet contribution to the entropy.

In contrast, the free-energy comparison leads to a different indication. 
Since the axially symmetric vectorized black holes have a higher Hawking temperature than the corresponding Schwarzschild black holes, their free energy is lower, as shown in Fig.~\ref{fig2}. 
Thus the entropy and free-energy criteria do not point in the same direction, and a complete thermodynamic analysis is left for future work.

%From a thermodynamic point of view, this is precisely what seems to happen, as seen in Fig.~\ref{fig2}. 
%The free energy $F$ of these magnetic black holes is indeed lower than that of the Schwarzschild black holes.

%The scaled entropy $4S_{\rm H}/(16\pi M^2)$ of these black holes, on the other hand, is almost degenerate with the Schwarzschild one, as seen in Fig.~\ref{fig3}(b).
%This is interesting since it shows that the lower area contribution is almost exactly canceled by the additional term from the GB term.

%The figure further shows the scaled Hawking temperature $8 \pi T_{\rm H} M$, scaled horizon area $A_{\rm H}/(16\pi M^2)$, scaled equatorial horizon radius $R_e/(2M)$, and the equatorial-to-polar horizon radius ratio $R_e/R_p$, all versus the scaled coupling.
%Since $R_e/R_p<1$ the black hole horizon of these vectorized black holes is prolate.

The figure further shows the scaled Hawking temperature $8 \pi T_{\rm H} M$, the equatorial-to-polar horizon radius ratio $R_e/R_p$, the scaled equatorial horizon radius $R_e/(2M)$, and the scaled horizon area $A_{\rm H}/(16\pi M^2)$ versus the scaled coupling.
All quantities are seen to vary monotonically with the coupling.
%Like the entropy, 
Interestingly, the temperature of the axially symmetric vectorized black holes is higher than the corresponding Schwarzschild temperature.
In this sense, these black holes are hotter.
This is also the reason that their free energy is lower than the Schwarzschild free energy, while their entropy is almost the same.
%In contrast, the horizon area and the equatorial horizon radius are smaller.
The deformation of the black hole horizon is quantified by the equatorial-to-polar horizon radius ratio.
Since $R_e/R_p < 1$ these vectorized black holes have prolate deformation.

\begin{figure}[h!]
\begin{center}
%\vspace{-0.5cm}
\mbox{
\hspace*{-0.5cm}\includegraphics[height=.25\textheight, angle =0]{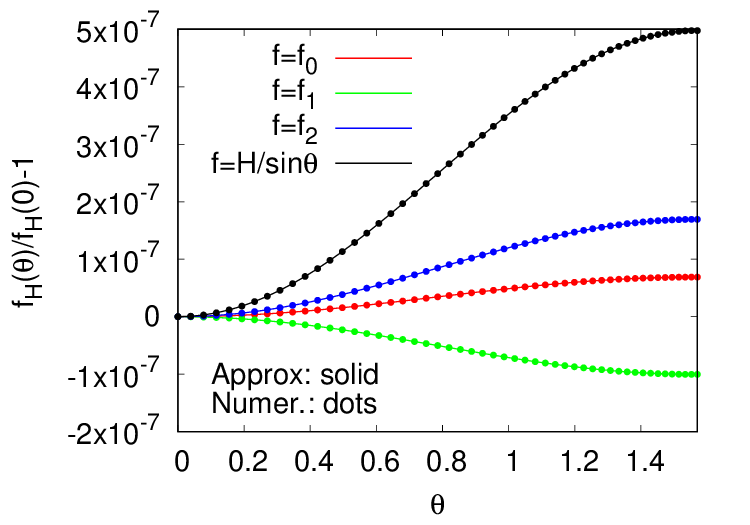}
}
\end{center}
\vspace{-0.5cm}
\caption{Static axially symmetric EvGB black holes:
Approximation Eq.~(\ref{xpand_LegPol}) (solid) and the numerical values (dots) at the horizon are shown as function of $\theta$ for $\lambda/M^2\approx (\lambda/M^2)_{\rm bif}$.
}
\label{fig4}
\end{figure}

Near the bifurcation with the Schwarzschild black hole the EvGB black holes can be approximated in terms of Legendre Polynomials $P_n(\cos\theta)$
\begin{equation}
f(r,\theta) = f^{(0)}(r) + f^{(2)}(r) P_2(\cos\theta) + \cdots \ ,
\label{xpand_LegPol}
\end{equation}
where $f = f_0$, $f_1$, $f_2$, and $H/\sin\theta$.
This is demonstrated in Fig.~\ref{fig4} for $\lambda/M^2\approx (\lambda/M^2)_{\rm bif}$. 
In order to emphasize the $\theta$-dependence, we show the normalized functions at the horizon $f_H(\theta)/f_H(0)-1$ (solid).
Also shown are the numerical values (dots).

The values of $(\lambda/M^2)_{\rm bif}$ can also be determined from lowest order perturbation theory.
In this case, the metric reduces to the Schwarzschild solution, and the vector field equations yield homogeneous linear ordinary differential equations (ODEs).
For the spherically symmetric EvGB black holes, the vector field is parametrized by $A = \gamma(r) dt$, where the function $\gamma(r)$ satisfies the ODE
\begin{equation}
\gamma^{''} 
      +\frac{2}{r}\frac{2+ (r/r_{\rm H})^2}{(r/r_{\rm H})^2-1}\, \gamma'
      -\frac{6}{r^2}\frac{1}{(r/r_{\rm H})^2-1}\, \gamma  
      +\frac{96 \lambda}{r_{\rm H}^4}\frac{(r/r_{\rm H})^2}{(1+r/r_{\rm H})^8} \, \gamma
 = 0 \ ,
\label{pertQ_ode}
\end{equation}
with boundary conditions $r_{\rm H}\gamma'(r_{\rm H})-\gamma(r_{\rm H})=0$ and $\gamma(\infty) = 0$.

Similarly, for the static axially symmetric EvGB black holes, the vector field is parametrized by $A = \gamma(r) \sin^2\theta d\varphi$, where the function $\gamma(r)$ satisfies the ODE
\begin{equation}
\gamma^{''} 
      -\frac{2}{r}\frac{1-2 r/r_{\rm H}}{(r/r_{\rm H})^2-1}\, \gamma'
      -\frac{2}{r^2}\, \gamma  
      +\frac{96 \lambda}{r_{\rm H}^4}\frac{(r/r_{\rm H})^2}{(1+r/r_{\rm H})^8} \, \gamma
 = 0 \ ,
\label{pertP2_ode}
\end{equation}
with boundary conditions $\gamma'(r_{\rm H})=0$ and $\gamma(\infty) = 0$.

Equations (\ref{pertQ_ode}) and (\ref{pertP2_ode}) have non-trivial solutions only for discrete values of $\lambda/M^2$, which increase with the number of nodes $n$ of $\gamma(r)$. 
For the lowest value, corresponding to node number $n=0$, we find for the spherically symmetric case $\lambda/M^2 = 4.6817$ and for the axially symmetric case $\lambda/M^2 = 3.1759$, in agreement with $(\lambda/M^2)_{\rm bif}$.

\begin{figure}[t!]
\begin{center}
%\vspace{-0.5cm}
\mbox{
(a)\hspace*{-0.5cm}\includegraphics[height=.25\textheight, angle =0]{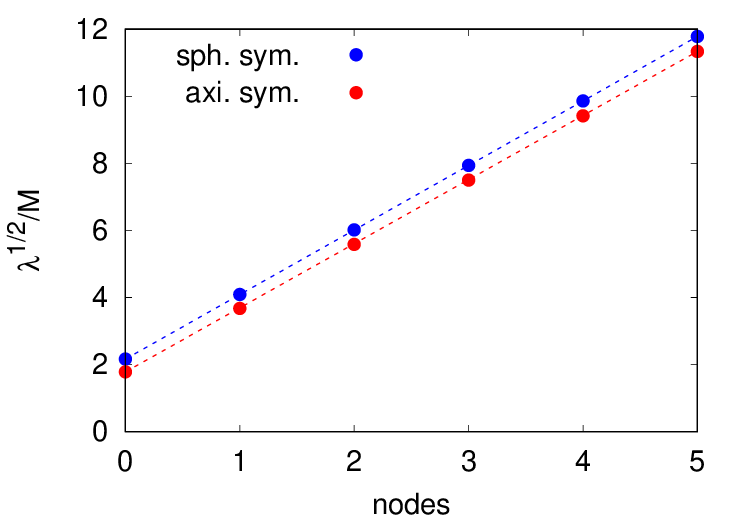}}
\mbox{
(b)\hspace*{-0.5cm}\includegraphics[height=.25\textheight, angle =0]{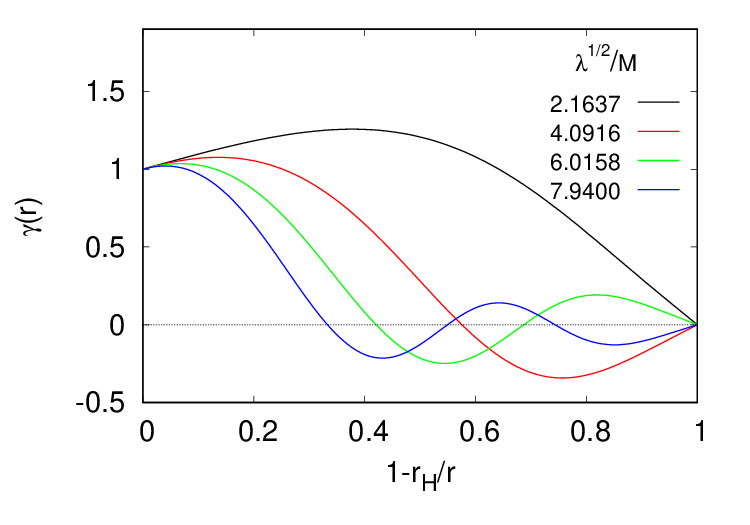}
(c)\hspace*{-0.5cm}\includegraphics[height=.25\textheight, angle =0]{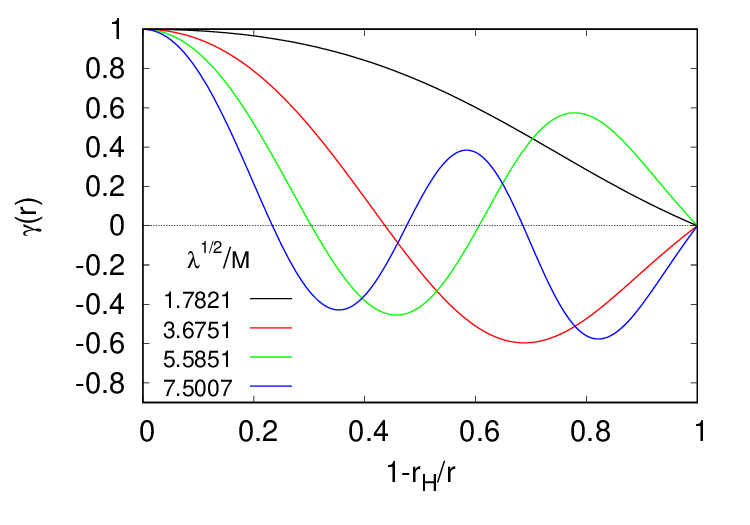}
}
\end{center}
\vspace{-0.5cm}
\caption{Perturbative solutions:
(a) Discrete values $\sqrt{\lambda}/M$ vs node number $n$, $n=0, \dots, 5$ and linear approximations (dotted).
(b) Modes $\gamma$ for the spherically symmetric case vs compactified coordinate $x=1-r_H/r$ for node numbers $n=0,\dots,3$.
(c) Same as (b) for the static axially symmetric case.}
\label{fig5}
\end{figure}

The discrete values $\sqrt{\lambda}/M$ are shown in Fig.~\ref{fig5}(a) for node numbers $n=0, \dots, 5$.
We observe that $\sqrt{\lambda}/M$ can be well approximated by straight lines with slopes $1.925$ and $1.911$ for the spherically symmetric and axially symmetric cases, respectively. 

The modes $\gamma(r)$ with node numbers $n=0, \dots, 3$ are shown in Fig.~\ref{fig5} as function of the compactified coordinate $x=1-r_{\rm H}/r $ for the static spherically symmetric case (b) and the static axially symmetric case (c) together with the corresponding values of $\sqrt{\lambda}/M$.

\subsection{Stationary rotating black holes} 

\begin{figure}[t!]
\begin{center}
%\vspace{-0.5cm}
\mbox{
(a)\hspace*{-0.5cm}\includegraphics[height=.25\textheight, angle =0]{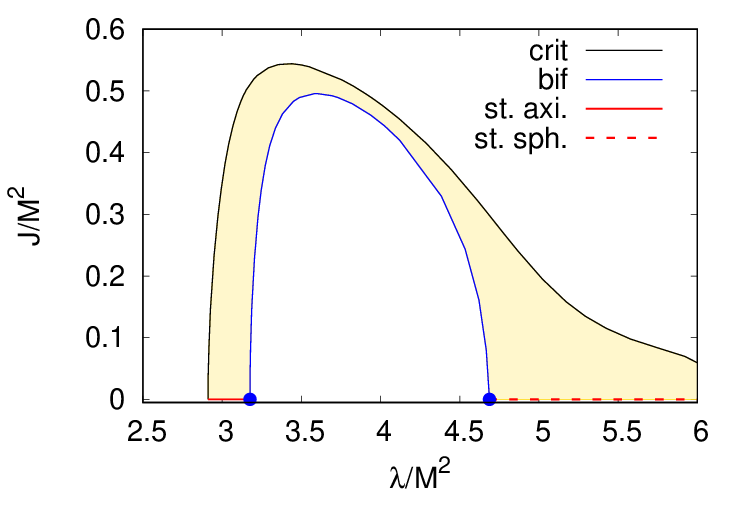}
(b)\hspace*{-0.5cm}\includegraphics[height=.25\textheight, angle =0]{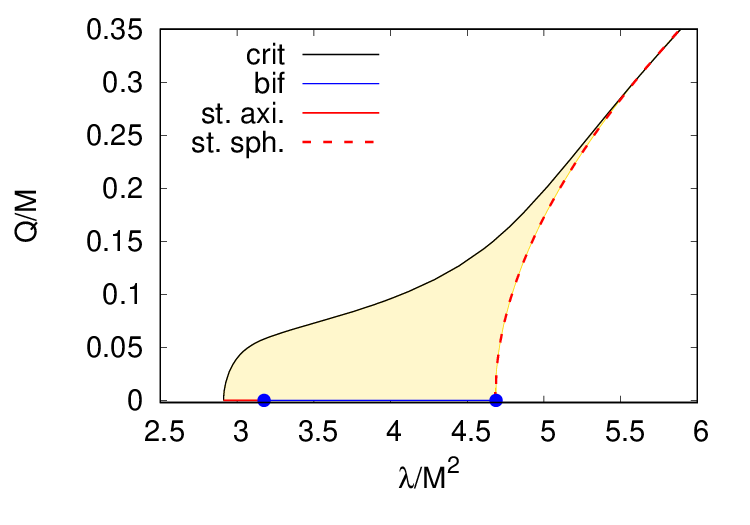}
}
\mbox{
(c)\hspace*{-0.5cm}\includegraphics[height=.25\textheight, angle =0]{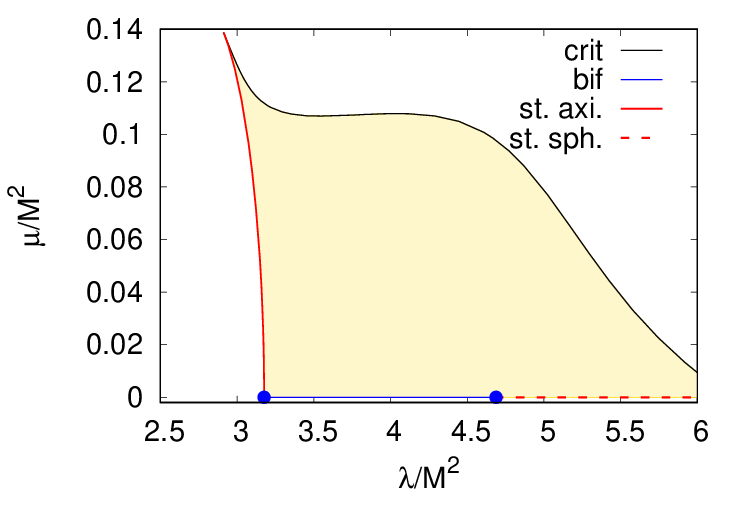}
(d)\hspace*{-0.5cm}\includegraphics[height=.25\textheight, angle =0]{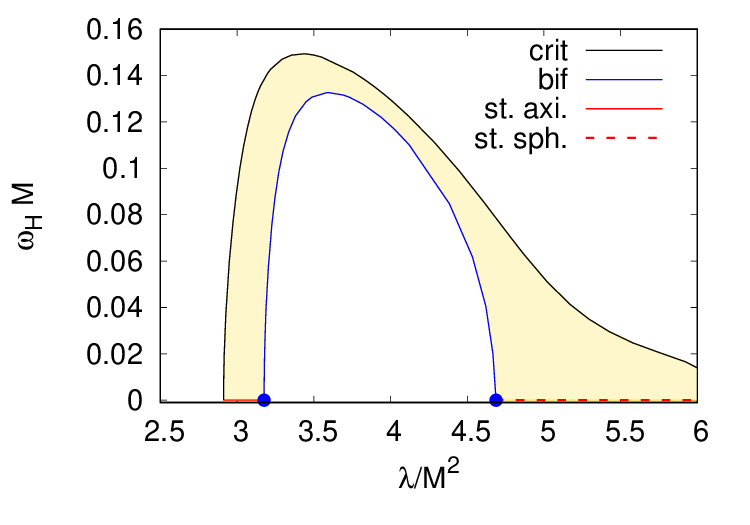}
}
\end{center}
\vspace{-0.5cm}
\caption{Rotating EvGB black holes:
(a) scaled angular momentum $J/M^2$ vs scaled coupling $\lambda/M^2$;
(b) scaled electric charge $Q/M$ vs scaled coupling; 
(c) scaled magnetic dipole moment $\mu/M^2$ vs scaled coupling;
(d) scaled horizon angular velocity $\omega_H M$ vs scaled coupling. 
}
\label{fig6}
\end{figure}

We now turn to the rotating generalizations of the above vectorized black holes.
Surprisingly, both types of static black holes possess a connected domain of existence, when rotation is included.
This is illustrated in Fig.~\ref{fig6}, where we show the scaled angular momentum $J/M^2$ (a), the scaled electric charge $Q/M$ (b), the scaled magnetic dipole moment $\mu/M^2$ (c), and the scaled horizon angular velocity $\omega_H M$ (d) versus the scaled coupling $\lambda/M^2$.
The domains of existence (yellow) of these quantities are delimited by the static axially symmetric solutions (solid red) with zero electric charge, the static spherically symmetric solutions (dashed red) with zero magnetic dipole moment, the bifurcation curves (blue) with both zero electric charge and zero magnetic dipole moment, and the critical solutions (black).

We note that static spherically symmetric black holes exist for arbitrarily large coupling parameter $\lambda$. 
Therefore we expect that rotating generalizations also exist for arbitrarily large $\lambda$.
However, we found that for large coupling parameter numerics becomes increasingly challenging. 
Therefore we restrict to moderate values of  $\lambda$.

\begin{figure}[t!]
\begin{center}
%\vspace{-0.5cm}
\mbox{
(a)\hspace*{-0.5cm}\includegraphics[height=.25\textheight, angle =0]{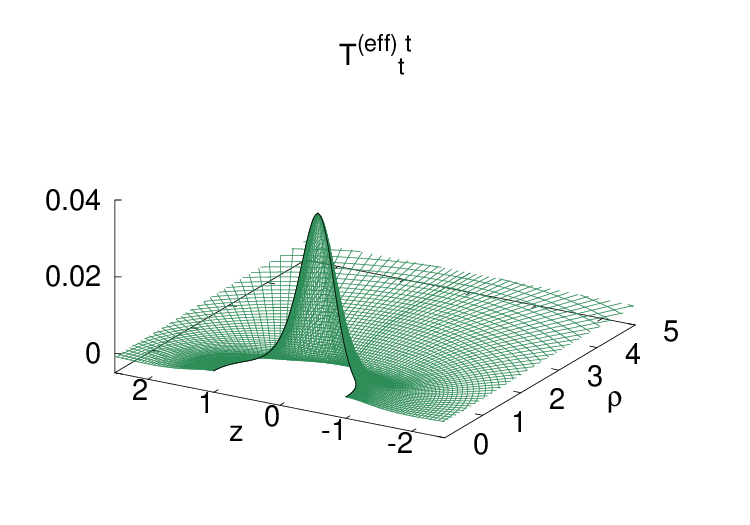}
(b)\hspace*{-0.5cm}\includegraphics[height=.25\textheight, angle =0]{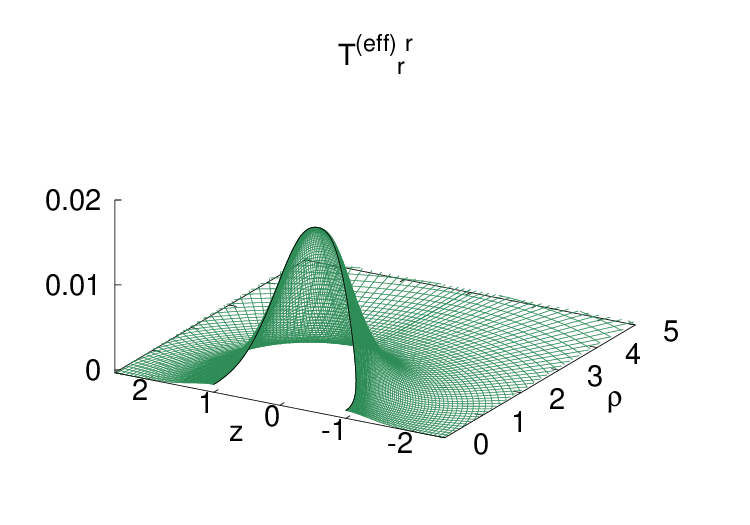}
}
\mbox{
(c)\hspace*{-0.5cm}\includegraphics[height=.25\textheight, angle =0]{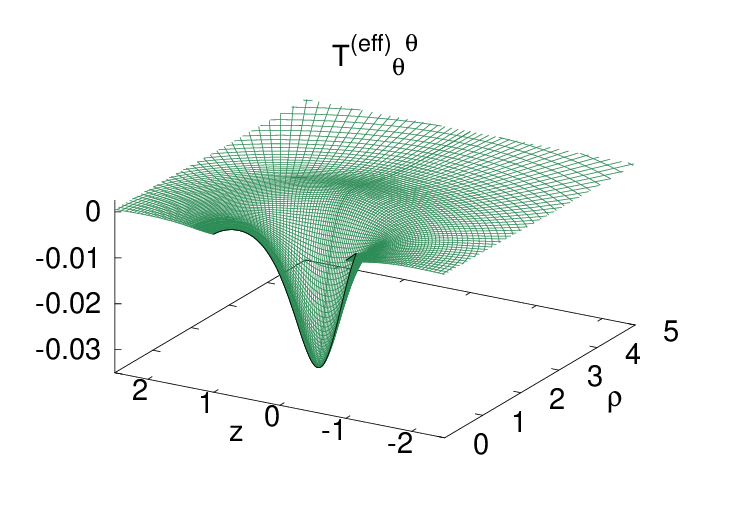}
(d)\hspace*{-0.5cm}\includegraphics[height=.25\textheight, angle =0]{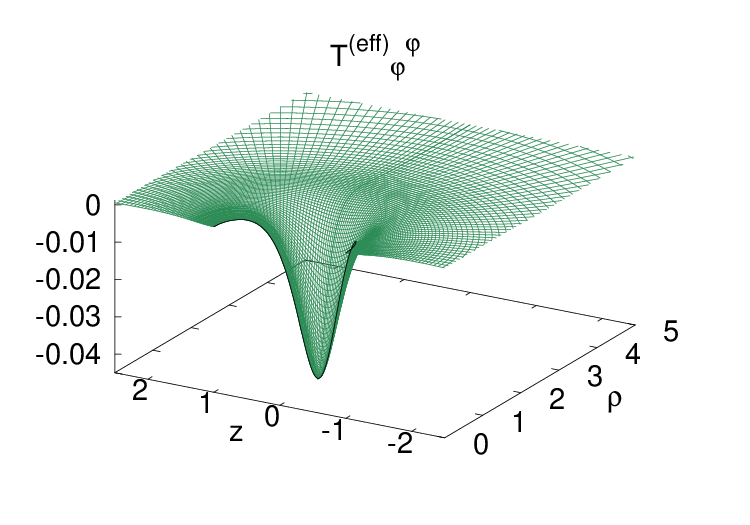}
}
\mbox{
(e)\hspace*{-0.5cm}\includegraphics[height=.25\textheight, angle =0]{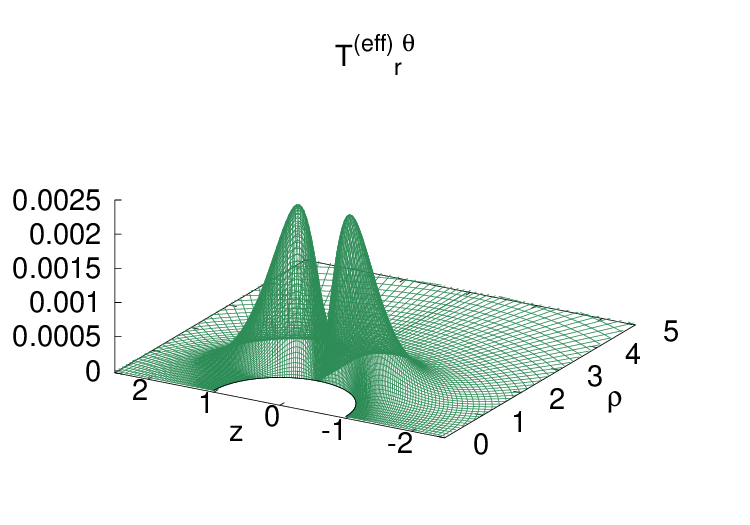}
(f)\hspace*{-0.5cm}\includegraphics[height=.25\textheight, angle =0]{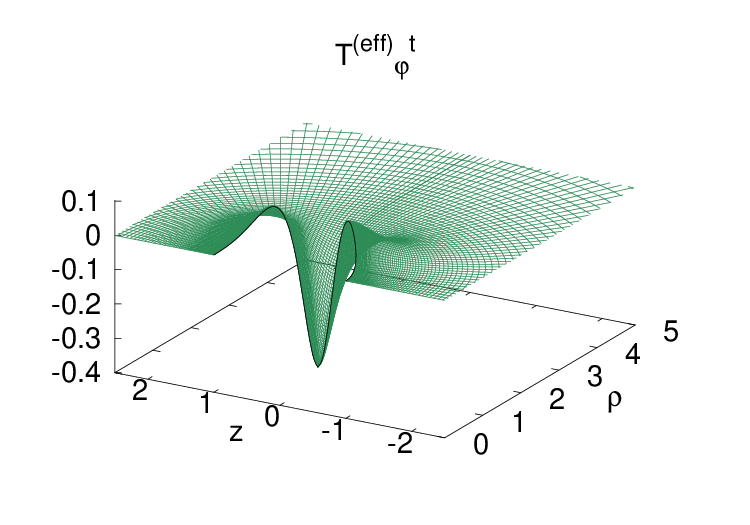}
}
\end{center}
\vspace{-0.5cm}
\caption{Rotating EvGB black holes:
effective stress-energy tensor components $T_{\ \ \ \ t}^{\text{(eff)}\,t}$ (a), $T_{\ \ \ \ r}^{\text{(eff)}\,r}$ (b), $T_{\ \ \ \ \theta}^{\text{(eff)}\,\theta}$ (c), $T_{\ \ \ \ \varphi}^{\text{(eff)}\,\varphi}$ (d), $T_{\ \ \ \ r}^{\text{(eff)}\,\theta}$ (e), and $T_{\ \ \ \ \varphi}^{\text{(eff)}\,t}$ (f) for $J/M^2 = 0.5211$, $\lambda/M^2 =  3.5815$.
}
\label{fig7}
\end{figure}

Figure \ref{fig6}(a) shows that %for small angular momenta, 
the domain of existence (including boundaries) of the rotating vectorized black holes is not simply connected, as becomes evident when reversing the sense of rotation, i.e., $J \to -J$.  
The maximal angular momentum of Kerr black holes, where vectorization arises, is given by $J/M^2=0.496$ \footnote{We recall that in the case of scalarization, $J/M^2=0.5$ marks the minimal angular momentum of Kerr black holes, where spin induced scalarized black holes arise \cite{Dima:2020yac,Hod:2020jjy,Doneva:2020nbb,Herdeiro:2020wei,Berti:2020kgk}.}.
The angular momentum of these vectorized rotating black holes is limited to $J/M^2 \le 0.544$.
Thus, observations of black holes with higher spin would rule out their vectorization in the model 
unless black hole solutions with higher angular momenta exist in further sectors of the model.
We exhibit components of the effective stress-energy tensor $T_{\ \ \ \ \mu}^{\text{(eff)}\,\nu}$ in Fig.~\ref{fig7} for a rotating black hole solution with $J/M^2 = 0.5211$ and $\lambda/M^2 =  3.5815$, close to the maximal value of the dimensionless angular momentum.

The scaled electric charge $Q/M$ is shown in Figure \ref{fig6}(b).
We note that rotation induces a small, finite electric charge for the set of rotating black holes that emerge from the static axially symmetric ones.
For those rotating black holes that emerge from the static spherically symmetric black holes, the electric charge does not change much due to rotation, but a magnetic dipole moment is induced.
The magnetic dipole moment $\mu/M^2$ is seen in Figure \ref{fig6}(c).
It features a large, compact domain of existence.
The horizon angular velocity $\omega_H M$, on the other hand, possesses a domain of existence resembling that of the angular momentum.

It is also interesting to consider the gyromagnetic ratio $g$, defined by $\mu = g\,\frac{QJ}{2M}$. 
For Kerr-Newman black holes one obtains $g=2$. 
For the rotating vectorized black holes, however, $g$ is not constant over the domain of existence. 
In particular, it can deviate substantially from the Kerr-Newman value, reflecting the non-electromagnetic nature of the vector field and the non-linear coupling to the Gauss-Bonnet invariant.

\begin{figure}[t!]
\begin{center}
%\vspace{-0.5cm}
\mbox{
(a)\hspace*{-0.5cm}\includegraphics[height=.25\textheight, angle =0]{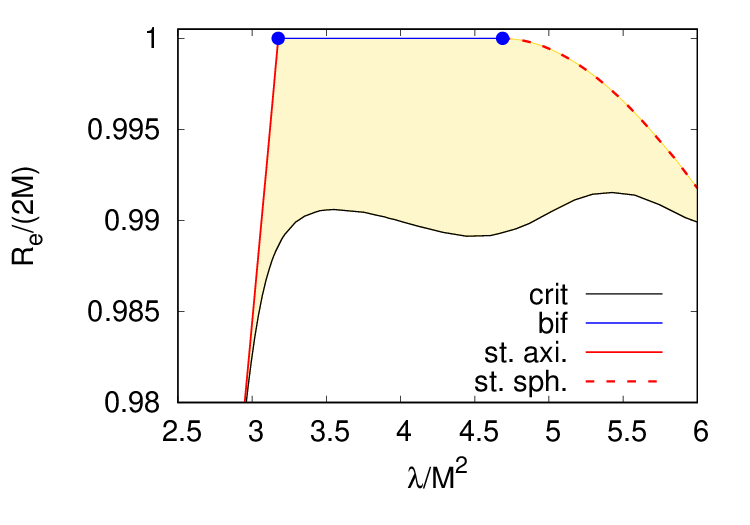}
(b)\hspace*{-0.5cm}\includegraphics[height=.25\textheight, angle =0]{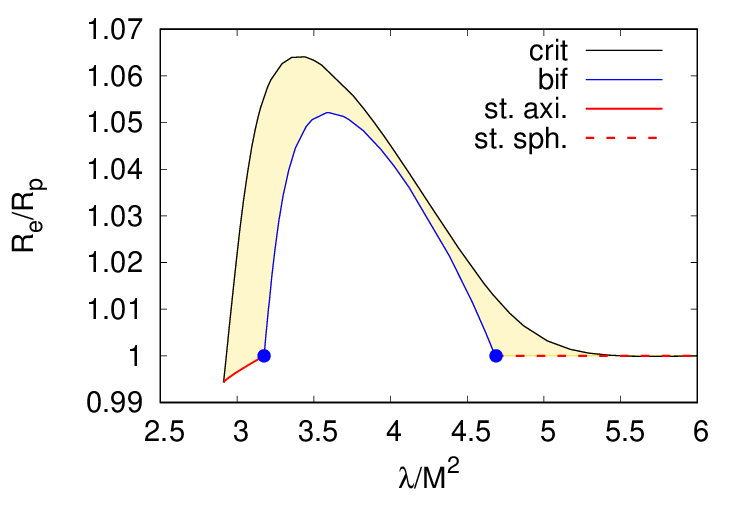}
}
\mbox{
(c)\hspace*{-0.5cm}\includegraphics[height=.25\textheight, angle =0]{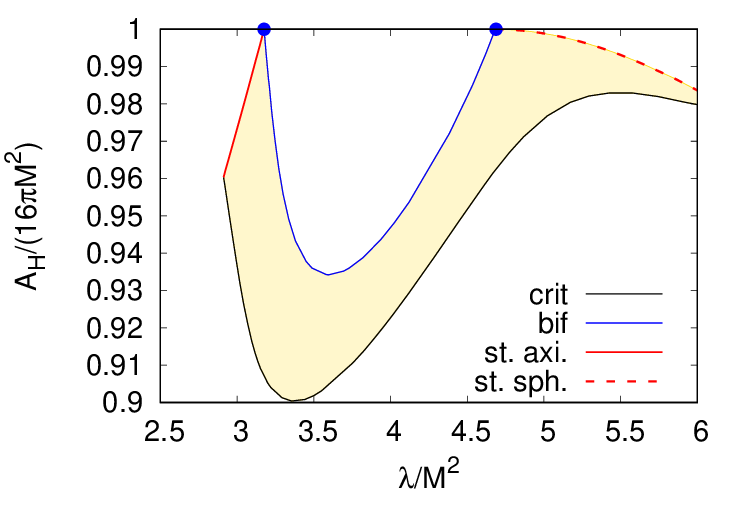}
(d)\hspace*{-0.5cm}\includegraphics[height=.25\textheight, angle =0]{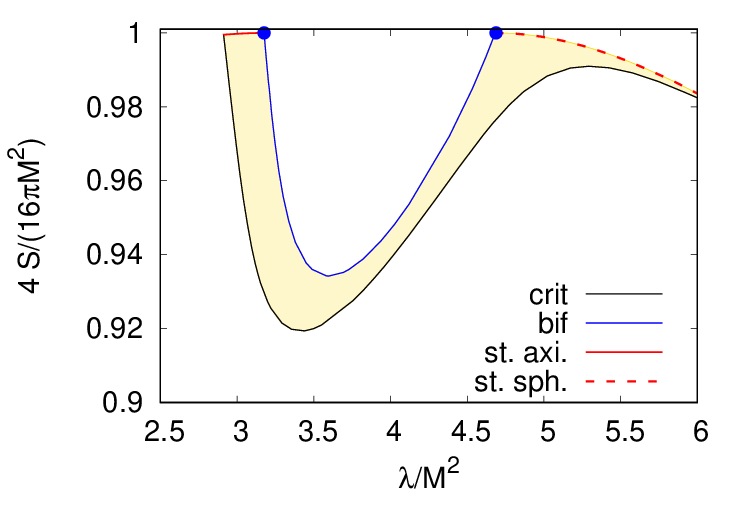}
}
\end{center}
\vspace{-0.5cm}
\caption{Rotating EvGB black holes:
(a) scaled equatorial horizon radius $R_e/(2M)$ vs scaled coupling $\lambda/M^2$;
(b) equatorial-to-polar horizon radius ratio $R_e/R_p$ vs scaled coupling;
(c) scaled horizon area $A_{\rm H}/(16\pi M^2)$ vs scaled coupling;
(d) scaled entropy $4 S_{\rm H}/(16\pi M^2)$ vs scaled coupling. 
}
\label{fig8}
\end{figure}

\begin{figure}[t!]
\begin{center}
%\vspace{-0.5cm}
\mbox{
(a)\hspace*{-0.5cm}\includegraphics[height=.25\textheight, angle =0]{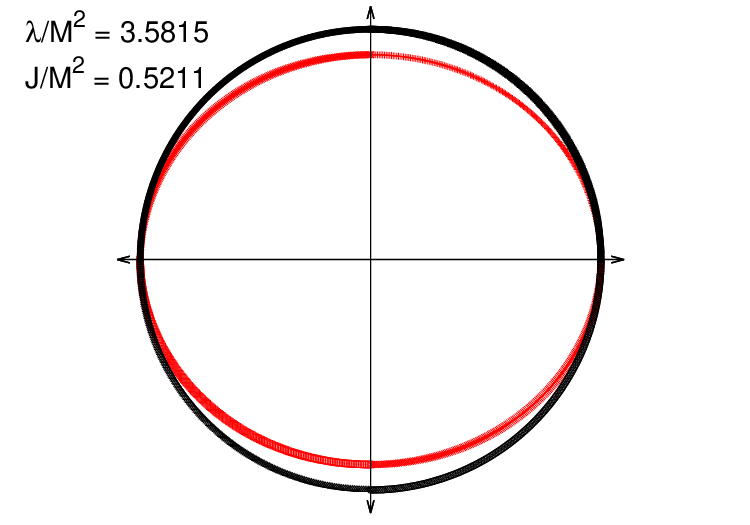}
(b)\hspace*{-0.5cm}{\vspace*{-1.5cm}\includegraphics[height=.25\textheight, angle =0]{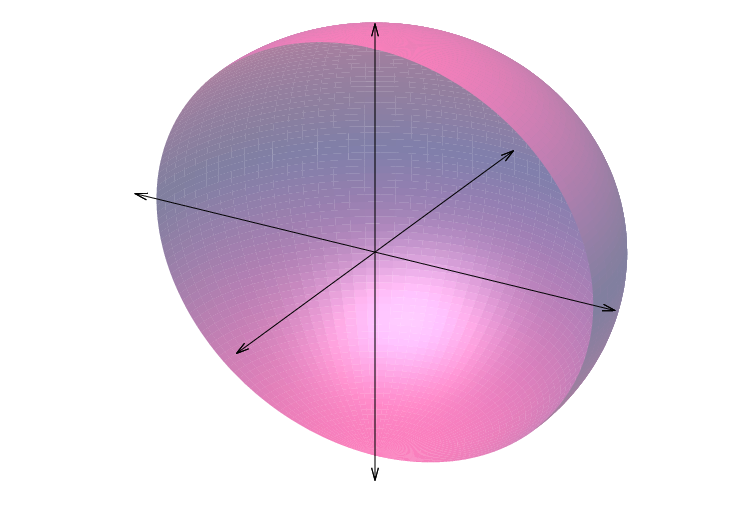}}
}
\end{center}
\vspace{-0.5cm}
\caption{Rotating EvGB black holes:
horizon embedding 2-dimensional (a), and 3-dimensional (b) for $J/M^2 = 0.5211$, $\lambda/M^2 =  3.5815$.
}
\label{fig9}
\end{figure}

We now turn to the horizon geometry.
Figure \ref{fig8}(a) exhibits the scaled equatorial horizon radius $R_e/(2M)$ versus the scaled coupling $\lambda/M^2$.
For the Kerr black holes this ratio remains constant, $R_e=2M$.
But the vectorized black holes have a smaller ratio, since the vector fields contribute to the mass outside the horizon.
The equatorial-to-polar horizon radius ratio $R_e/R_p$ of the rotating black holes shown in Fig.~\ref{fig8}(b) reveals a predominantly oblate horizon deformation, since rotation causes a centrifugal flattening.
So, already at $J/M^2 \approx 0.16$ %\jk{???} 
the rotating generalizations of the prolate static black holes become oblate.
Curiously, however, in the slow rotation sector of the originally spherically symmetric black holes, the horizon becomes prolate in spite of rotation.
Embedding diagrams of the horizon of the rapidly rotating black holes of Fig.~\ref{fig7} are shown in Fig.~\ref{fig9}.

Figure \ref{fig8}(c) presents the scaled horizon area $A_{\rm H}/(16\pi M^2)$ versus scaled coupling, which is always smaller than the area of the associated Schwarzschild or Kerr black holes.
The scaled entropy $4 S_{\rm H}/(16\pi M^2)$ versus the scaled coupling is shown in Fig.~\ref{fig8}(d).
The additional term from the Gauss-Bonnet coupling increases the entropy considerably with respect to the area.
But the presence of rotation does not increase the entropy of the vectorized black holes above the Kerr entropy.
%But in spite of the almost equivalence of deformed static black holes and Schwarzschild black holes, the presence of rotation does not make the vectorized black holes thermodynamically preferred (in a micro-canonical ensemble). 

\begin{figure}[b!]
\begin{center}
%\vspace{-0.5cm}
\mbox{
(a)\hspace*{-0.5cm}\includegraphics[height=.25\textheight, angle =0]{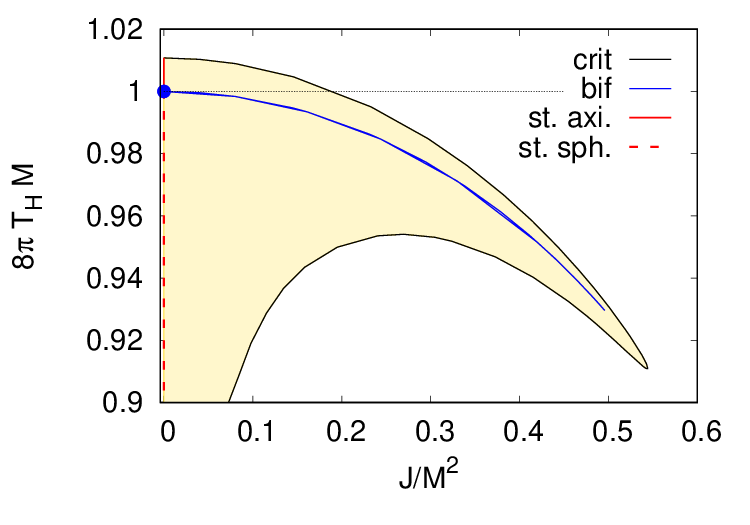}
(b)\hspace*{-0.5cm}\includegraphics[height=.25\textheight, angle =0]{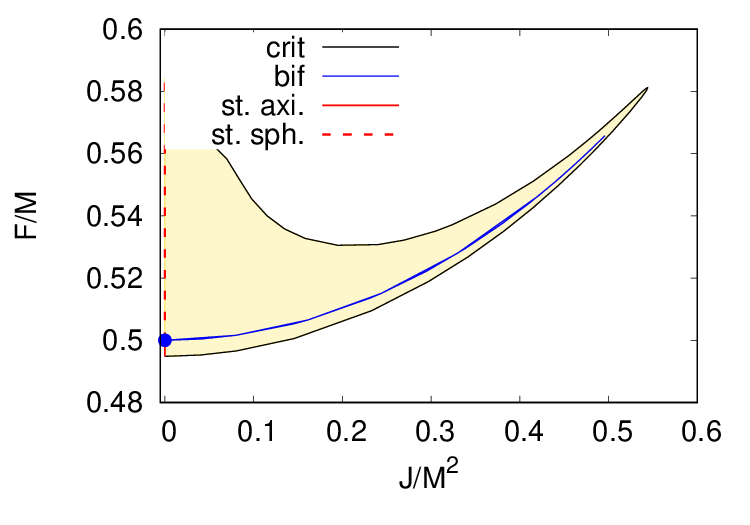}
}
%\mbox{
%(c)\hspace*{-0.5cm}\includegraphics[height=.25\textheight, angle =0]{rdomx_Fm_vs_lm2.eps}
%(d)\hspace*{-0.5cm}\includegraphics[height=.25\textheight, angle =0]{rdomx_Fm_vs_Tm.eps}
%}
\end{center}
\vspace{-0.5cm}
\caption{Rotating EvGB black holes:
(a) scaled Hawking temperature $8 \pi T_{\rm H} M$ vs scaled angular momentum $J/M^2$;
%scaled coupling $\lambda/M^2$;
(b) scaled free energy $F/M$ vs scaled angular momentum. % $J/M^2$. 
}
\label{fig10}
\end{figure}

We exhibit in Fig.~\ref{fig10} the scaled Hawking temperature $8 \pi T_{\rm H} M$ (a) and the scaled free energy $F/M$ (b) versus the scaled angular momentum $J/M^2$. 
Thus, the rotating generalizations of the deformed static black holes are hotter than the corresponding Kerr black holes.
As in the static case, this entails a lower free energy for a given $J/M^2$ for this set of rotating vectorized black holes, as seen in the figure.

\section{Conclusion}

We have considered spontaneously vectorized black holes in an Einstein-vector-Gauss-Bonnet theory with a massless vector field $A_\mu$ and the coupling function $\lambda A_\mu A^\mu$.
Static, spherically symmetric vectorized black holes of this theory have been studied before and shown to exist above a minimal value of the coupling $\lambda$. 
Here we have shown that, in addition to these electrically charged black holes, there are uncharged static black holes that possess a magnetic dipole moment and a prolate deformation.
Clearly, Israel's theorem does not hold for these vectorized black holes.

These magnetic black holes bifurcate from the Schwarzschild solutions at a lower value of the coupling constant than the electric black holes and exist in a finite interval of the scaled coupling constant, $\lambda_\text{cr}/M^2 < \lambda/M^2 < \lambda_\text{bif}/M^2$.
Interestingly, their entropy is almost degenerate with the entropy of the Schwarzschild black holes.
While their horizon area is lower, the additional contribution to the entropy from the GB term almost cancels the difference in area.
Since these magnetic black holes feature a higher temperature than the Schwarzschild black holes, their free energy is smaller.

When the static black holes are set into rotation, the electric black holes acquire a magnetic dipole moment and the magnetic black holes an electric charge.
For slow rotation, their domains of existence are still distinct, but for more rapid rotation, their domains merge.
The full domain of existence including the boundaries is thus not simply connected when the angular momentum is considered as a function of the coupling constant.
The inner boundary is provided by the bifurcation line from the GR solutions, and the outer boundary corresponds to the critical solutions with a maximal dimensionless angular momentum of $(J/M^2)_\text{max}=0.544$. %\jk{???}

This value is far below the Kerr bound, where extremal black holes occur.
We thus do not obtain extremal vectorized black holes.
This is also seen in the horizon temperature, which remains rather large in the presence of rotation.
The entropy of the vectorized black holes never exceeds the entropy of the GR black holes.
But the vectorized black holes possess, in part, a lower free energy than the GR black holes. 
A full analysis of the thermodynamic properties of the vectorized black holes is, however, beyond the scope of the current study and is deferred to the future.

We finally comment on a limitation of the present model. 
It has been argued in related vectorization theories that the tachyonic mechanism responsible for the growth of the vector field is accompanied by the appearance of ghost-like degrees of freedom, or equivalently by a loss of hyperbolicity of the vector-field evolution equations \cite{Garcia-Saenz:2021uyv,Silva:2021jya,Clough:2022ygm,Coates:2022qia,Pizzuti:2023eyt,Chen:2024hkm}.
This represents a dynamical obstruction to the physical viability of such theories, which is not visible at the level of stationary solutions alone. 
The present work should therefore be understood as a construction and characterization of stationary black hole solutions of the EvGB field equations, including their global, horizon, and thermodynamic properties. 
The question whether these solutions can arise in a well-posed dynamical evolution is not addressed here and is strongly constrained by the pathologies identified in the above references.

Independently of this issue, quasinormal modes (QNMs) would be of interest for characterizing perturbations of the stationary solutions and for comparison with other modified-gravity black holes.
%Another important aspect left for future study is the investigation of the \jk{linear stability of these vectorized solutions. 
%In particular, vectorized black holes with vector hair may be affected by ghost instabilities, as pointed out in related vector-tensor theories \cite{Garcia-Saenz:2021uyv,Silva:2021jya,Clough:2022ygm,Coates:2022qia,Pizzuti:2023eyt,Chen:2024hkm}.
%Such instabilities would clearly be relevant for the astrophysical viability of these solutions. 
%Quasinormal modes (QNMs) 
%provide one way to address this question} and 
QNMs are %also 
relevant for modeling the ringdown after the merger of black holes (see e.g.~\cite{Berti:2025hly}). 
In recent years, numerical methods to obtain the QNMs of rapidly rotating black holes in modified gravity theories with additional fields have been developed \cite{Berti:2025hly,Blazquez-Salcedo:2023hwg,Chung:2024ira,Chung:2024vaf,Blazquez-Salcedo:2024oek}.
These methods may also be useful for studying perturbations of %seem suitable for extracting the QNMs also for 
rotating EvGB black holes.

%Another important aspect left for future study is the investigation of the quasinormal modes (QNMs) of these vectorized solutions.
%QNMs are relevant for modeling the ringdown after the merger of black holes (see e.g.~\cite{Berti:2025hly}).
%Moreover, possible (linear) instabilities of black holes can be uncovered this way.

\end{document}